\documentclass[twocolumn,aps,amsmath,amssymb,floatfix,prb]{revtex4}
\usepackage{graphicx}
\usepackage{dcolumn}
\usepackage{bm}
\usepackage{amsfonts}
\usepackage{graphicx}
\usepackage{dcolumn}
\usepackage{bm}
\usepackage{amsmath}
\usepackage{amssymb}

\newcommand{\vk}{{\bf k}}
\newcommand{\ve}{{\varepsilon}}
\newcommand{\vek}{{\varepsilon_k}}

\newcommand{\vq}{{\bf q}}

\begin{document}

\title{Impurity scattering induced carrier transport in twisted bilayer graphene}
\author{E. H. Hwang$^{1,2}$ and S. Das Sarma$^{1,3}$}
\affiliation{
$^1$Condensed Matter Theory Center, 
Department of Physics, University of Maryland, College Park,
Maryland  20742-4111 \\
$^2$SKKU Advanced Institute of Nanotechnology and Department of
Nano Technology, Sungkyunkwan  University, Suwon, 16419, Korea \\
$^3$Joint Quantum Institute, University of Maryland, College Park,
Maryland  20742-4111 
}



\begin{abstract}
We theoretically calculate the impurity-scattering induced resistivity of twisted bilayer graphene at low twist angles where the graphene Fermi velocity is strongly suppressed.  We consider, as a function of carrier density, twist angle, and temperature, both long-ranged Coulomb scattering and short-ranged defect scattering within a Boltzmann theory relaxation time approach.  For experimentally relevant disorder, impurity scattering contributes a resistivity comparable to (much larger than) the phonon scattering contribution at high (low) temperatures.  Decreasing twist angle leads to larger resistivity, and in general, the resistivity increases (decreases) with increasing temperature (carrier density).  Inclusion of the van Hove singularity in the theory leads to a strong increase in the resistivity at higher densities, where the chemical potential is close to a van Hove singularity, leading to an apparent density-dependent plateau type structure in the resistivity, which has been observed in recent transport experiments.  We also show that the Matthissen's rule is strongly violated in twisted bilayer graphene at low twist angles.
\end{abstract}

\maketitle


\section{Introduction}

Electronic properties, particularly ohmic transport properties, of twisted bilayer graphene (TBLG) are of great current interest because of the seminal experimental findings by Cao et al.\cite{one,two} at MIT that TBLG has intriguing low-temperature density- and temperature-dependent transport behavior.  In particular, both superconducting and insulating ground states seem to exist in TBLG at various carrier densities and low twist angles.\cite{one,two,three,four} The resultant density-temperature-twist angle dependent TBLG phase diagram is rich and complex, and is being actively studied in many laboratories.  Although there are many proposed theories for TBLG ground states, there is no consensus yet on the nature of the superconducting (S) or insulating (I) ground states.
 
Our theoretical work is on electronic transport above the ground state, i.e., at elevated temperatures (much larger than the corresponding gap defining the S or I phase) where superconducting or insulating behavior is suppressed and the system behaves like an effective metal, as found experimentally.\cite{five,six} The issue we address is how disorder in TBLG samples affects the ohmic transport properties, assuming that the system can be considered to be an effective 2D metal with a finite carrier density.  Effect of phonon scattering on TBLG `metallic' transport has recently been considered in the literature\cite{seven}, so we focus on the effect of impurity scattering.  One motivation for our considering impurity scattering effects is the fact that disorder is known to be the most important resistive scattering source in regular (i.e., untwisted) graphene up to room temperatures because the typical electron-phonon coupling in regular graphene is weak.  
It has been argued in Ref.~[\onlinecite{seven}] that the strong suppression in the TBLG Fermi velocity at low twist angles leads to a giant enhancement in the effective electron-phonon coupling, causing  a large contribution to the phonon-induced temperature-dependent resistivity.  In the current work, we address the issue of the effect of TBLG Fermi velocity suppression on impurity scattering-induced graphene resistivity.  In addition, we investigate the extent to which the Matthiessen's rule applies to TBLG transport at finite temperatures by taking into account resistive scattering from both phonons and impurities.   We find that Matthiessen's rule is strongly violated in TBLG at low twist angles leading to the actual resistivity being 100\% (or more) larger than the sum of the individual impurity and phonon resistivities.

We provide the basic transport theory for calculating the resistivity arising from both electron-phonon coupling and electron-impurity scattering in Sec. II. We give detailed results for $\rho(T,n)$ for twisted bilayer graphene also in Sec. II, emphasizing
the failure of  Matthiessen's rule at low twist angles where the Fermi velocity is small. In Sec. III, we provide resistivity results with the inclusion of van Hove singularity in the electronic density of states. 
Finally, a brief conclusion is given in Sec. IV. 

\section{Theory and Results}

Detailed theories for impurity scattering effects in graphene are already available in the literature,\cite{eight,nine,ten} which would not be reproduced here since we use the standard theory\cite{eight,nine,ten} involving Boltzmann equation and relaxation time approximation within the leading order scattering approximation.  The main question in calculating the impurity resistivity is modeling the impurity scattering potential and the TBLG electronic structure.  Unfortunately, neither is well-established at this early developing stage of the subject, and in fact, even in regular untwisted graphene, settling the precise nature of disorder scattering limited resistivity took some time.\cite{eight,nine,ten,eleven,twelve,thirteen,fourteen,fifteen} 
 
The nature of the dominant impurity scattering in TBLG is not known, and the precise TBLG electronic structure is also unknown.  In particular, it is believed that there is some twist angle fluctuation related disorder in TBLG, but there is no available quantitative information on this disorder.  It is reasonable to assume that disorder effects existing in regular graphene, random charged impurities and point defects, are also present in TBLG since TBLGs are formed by two monolayers of regular graphene.
 Following the well-established disorder model in untwisted graphene, we assume that TBLG has two types of disorder:  long-range disorder arising from random charged impurities and short-range disorder arising neutral impurities and defects.  We subsume the unknown twist angle fluctuation disorder as contributing to the short-range disorder part of our model.  Our impurity model thus has two unknown independent parameters corresponding to the density of random charged impurities and the strength of the short range disorder. 
 
The dominant physics of the TBLG electronic structure affecting transport properties is the twist angle dependent flattening of the moir\'e superlattice bands operational in the system.  This band flattening leads to a strong suppression of the graphene Fermi velocity with decreasing twist angle.  Following Ref.~[\onlinecite{seven}], we include in the theory the band flattening effect through a modified Fermi velocity arising from the moir\'e superlattice structure of the system.  The dependence of the Fermi velocity on the twist angle is already given in Ref.~[\onlinecite{seven}], and not reproduced here. We show our results as a function of the TBLG Fermi velocity which we take as a variable -- the dependence of this reduced Fermi velocity on the twist angle follows the electronic structure model introduced in Ref.~[\onlinecite{sixteen}] and the corresponding Fermi velocity $v_F$ as a function of the twist angle is given in [\onlinecite{seven}].

More sophisticated electronic structure can be incorporated in the theory at the considerable price of all analytical transparency being lost since all matrix elements must be calculated numerically, which would be quite demanding for a finite temperature transport calculation of interest here.  Also, using complicated electronic structure for transport calculations may be an unnecessary overkill at this stage of development of the subject since the precise nature and the quantitative details of the underlying disorder in TBLG are unknown right now.  More importantly, the electronic structure of the TBLG moir\'e superlattice is far from being accurately known with considerable sample to sample variations.  These variations could arise from strain relaxation in the TBLG and/or from twist angle fluctuations or from other unknown sources.  Therefore, it makes sense in this early stage of the subject to use a minimal model for the TBLG electronic structure, which obviously is the incorporation of the flatband induced twist angle dependent Fermi velocity suppression in the theory.  

At higher carrier density, as the chemical potential approaches a van Hove singularity (vHS) associated with the moir\'e superlattice, we include the vHS effect in the transport calculation using a model density of states incorporating vHS effects qualitatively as described later in this paper in the next section.

First, we discuss the situation without considering vHS effects, which is valid at relatively low carrier densities ($n< 2 \times 10^{12}$ cm$^{-2}$) near the Dirac point.  For doping densities not too far from the Dirac point (for less than `quarter filling' either on the electron or the hole side), the TBLG Fermi level or chemical potential is well below any vHS, and neglecting the vHS effect is a valid approximation.  The resistivity $\rho=1/\sigma$, where the conductivity $\sigma$ is given by
\begin{equation}
\sigma(n,T) = \frac{e^2}{h} \frac{g v_F k_F}{2} \langle \tau \rangle,
\label{eq1_sigma}
\end{equation} 
where $g$ is the total degeneracy, $k_F$ is the Fermi wave vector, and $\langle \tau \rangle$ is the relaxation time averaged over energy. 
For impurity scattering we consider the screened long range Coulomb disorder and unscreened short range disorder. The impurity scattering of long range disorder is determined by the impurity charge density $n_i$, and the scattering of short range disorder is determined by the parameter $n_dV_0^2$, where $n_d$ is the impurity density and $V_0$ is the strength of the potential.  

In order to find the total resistivity at finite temperatures we have to calculate the energy averaged transport relaxation time $\langle \tau \rangle$ after adding the individual scattering rates due to impurities (i) and acoustic phonons (ph),
i.e., 
\begin{equation}
\tau_{\rm tot}^{-1} = \tau_i^{-1} + \tau_{\rm ph}^{-1}, 
\label{eq2_time}
\end{equation}
which deviates from adding the averaged individual scattering rate, i.e., $\langle \tau_{\rm tot}^{-1} \rangle = \langle \tau_{i}^{-1} \rangle  + \langle \tau_{\rm ph}^{-1} \rangle $.  
The other important temperature effect of scattering times in our model comes from temperature dependent screening in the screened long range disorder. We consider the temperature dependent screening (or dielectric function), i.e.,
\begin{equation}
\epsilon(q,T) = 1 - v(q) \Pi(q,T), 
\end{equation}
where $v(q) = 2\pi e^2/\kappa q$ is the electron-electron interaction with a background dielectric constant $\kappa$ and $\Pi(q,T)$ is the irreducible finite-temperature polarizability function.\cite{eight,nine} 
The main effect in TBLG is a reduction in the Fermi velocity as twisted angle is reduced. In the calculation we incorporate the renormalized Fermi velocity in the polarizability function.  
For phonon scattering, we follow Refs.~[{\onlinecite{seven,seventeen,eighteen}], which we do not reproduce here.  Note that there are a number of variables and system parameters determining the TBLG resistivity: carrier density (determining $k_F$, $E_F$, etc.), twist angle (determining electronic structure and particularly, $v_F$), disorder strength (characterized by the parameters for long- and short-range impurities), phonon scattering strength (which, following Ref.~[\onlinecite{seven}], we take to be the deformation potential coupling appropriate for TBLG). (See the appendix for the details of acoustic phonon induced resistivity.)

\begin{figure}[b]
\vspace{10pt}%
\includegraphics[width=1.\linewidth]{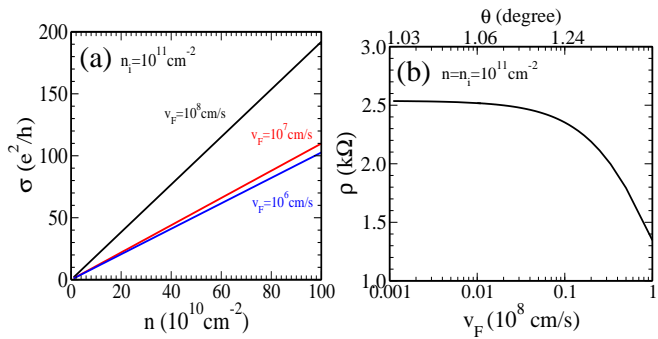}
\caption{
(a) Calculated conductivity as a function of carrier density for a charged impurity density ($n_i=10^{11}$cm$^{-2}$) and for three different Fermi velocities $v_F=1$, 0.1, 0.01$\times10^8$ cm/s. (b) Calculated resistivity as a function of Fermi velocity for a fixed carrier density and for an impurity density ($n=n_i=10^{11}$cm$^{-2}$).
The TBLG twist angles corresponding to the $v_F$-values in panel (a) are: $\theta=25^{\circ}, \; 1.24^{\circ}, \; 1.06^{\circ}$.
}
\label{fig:one}
\end{figure}

At zero temperature we have the asymptotic behavior of the conductivity as a function of Fermi velocity; for  long range disorder
\begin{eqnarray}
\sigma_0(v_F) & \sim & {\rm const.} \;\; {\rm for} \; v_F \rightarrow 0 \; ({\theta \rightarrow \rm magic \; angle)} \nonumber \\
                        & \sim & v_F^2 \;\;\;\; {\rm for} \; v_F \rightarrow \infty,
\end{eqnarray} 
and for short range disorder
\begin{equation}
\sigma_0(v_F)  \sim v_F^2 \;\; {\rm for \; all} \; v_F.
\end{equation}
For the long range disorder $v_F \rightarrow 0$ (i.e., $\theta \rightarrow $ magic angle) corresponds to the stron screening limit, and $v_F \rightarrow \infty$ the weak screening limit.
For a fixed Fermi velocity the density dependent conductivity becomes 
$\sigma(n) \sim n$ for long range disorder and  for short range disorder $\sigma (n)$ is independent of the carrier density.

\begin{figure}[t]
\vspace{10pt}%
\includegraphics[width=1.\linewidth]{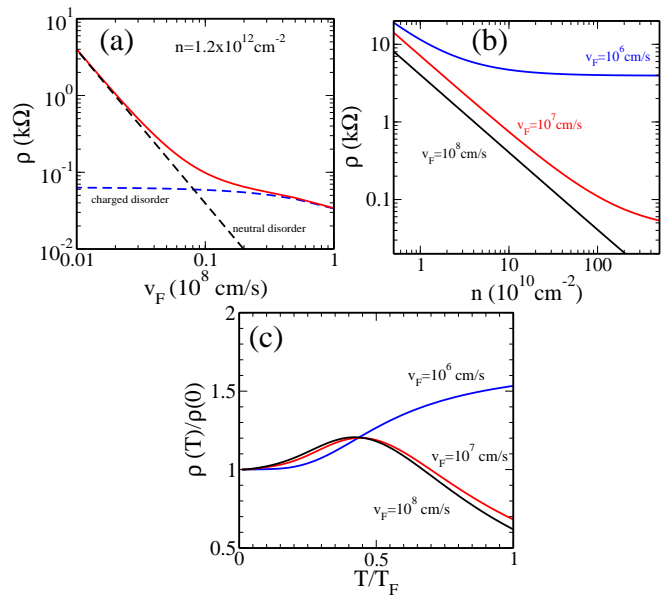}
\caption{ (a) Calculated resistivity with both long range charged disorder and short range neutral disorder as a function of $v_F$ for a fixed carrier density $n=1.2\times 10^{12}$cm$^{-2}$. Here the charged impurity density $n_i=3\times 10^{10}$ cm$^{-2}$ and the neutral disorder strength $n_dV_0^2 = 0.53$ (eV\AA)$^2$ are used. (b) The resistivity as a function of density for several $v_F$ with the same disorder parameters as (a).
(c) The temperature dependent resistivity, $\rho(T)/\rho(0)$ where $\rho(0)$ is the resistivity at $T=0$, as a function of $T/T_F$ for several $v_F$ with the same disorder parameters as (a).
Twist angles corresponding to $v_F=10^8,\; 10^7,\; 10^6$ cm/s are $\theta=25^{\circ}, \; 1.24^{\circ}, \; 1.06^{\circ}$, respectively.
}
\label{fig:two}
\end{figure}

Figure~\ref{fig:one} shows the calculated transport behavior of long range disorder. In Fig.~\ref{fig:one}(a) the calculated conductivity is shown as a function of carrier density with a charged impurity density 
and for three different fermi velocities $v_F=1$, 0.1, 0.01$\times10^8$ cm/s
corresponding to twist angle $\theta=25^{\circ}, \; 1.24^{\circ}, \; 1.06^{\circ}$, respectively.
As expected the conductivity increases linearly with carrier density for all Fermi velocities. In Fig.~1(b) the calculated resistivity as a function of Fermi velocity is shown at a fixed carrier density $n=10^{11}$ cm$^{-2}$. The resistivity arising from long range disorder saturates as $v_F \rightarrow 0$.

Figure~\ref{fig:two} shows the calculated resistivity with both long range charged disorder and short range neutral disorder.
In Figs.~\ref{fig:two}(a) and (b) the resistivity is shown as a function of Fermi velocity $v_F$ for a fixed carrier density $n=1.2\times 10^{12}cm^{-2}$ and as a function of carrier density for several $v_F$, respectively. The same disorder densities are used in the calculation, i.e., the charged impurity density $n_i=3\times 10^{10} cm^{-2}$ and the neutral disorder strength $n_dV_0^2 = 0.53\; eV^2 \AA^2$. As shown in the figures the long range charged (short range neutral) disorder dominates at large (small) Fermi velocities. In Fig.~\ref{fig:two}(c) the finite temperature resistivity normalized to the zero temperature resistivity, $\rho(T)/\rho(0)$, is shown as a function of $T/T_F$ for several $v_F$. In this calculation the phonon scattering is not included and the same disorder parameters as Fig.~\ref{fig:two}(a) are used. When the charged disorder scattering dominates the resistivity shows crossover (metallic to insulating) behavior
induced by screening\cite{ten,fourteen},
but it increases monotonically when the neutral disorder scattering dominates.

\begin{figure}[t]
\vspace{10pt}%
\includegraphics[width=1.\linewidth]{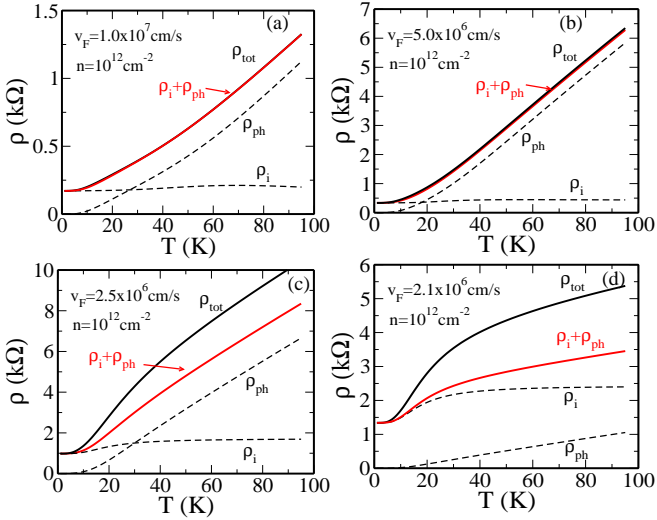}
\caption{
Failure of Matthiessen's rule at $v_{F} \sim v_{\rm ph}$. The calculated resistivities are shown for various Fermi velocities (twist angles); (a) $v_F=10^7$ cm/s ($\theta=1.24^{\circ}$), (b) $v_F=5\times 10^6$ cm/s ($\theta=1.14^{\circ}$), (c) $v_F=2.5\times 10^6$ cm/s ($\theta=1.09^{\circ}$), and (d) $v_F=2.1\times 10^6$ cm/s ($\theta=1.08^{\circ}$). The black lines indicate 
$\langle \tau_{\rm tot}^{-1} \rangle = \langle \tau_{\rm ph}^{-1} + \tau_{i}^{-1} \rangle $ and the red lines indicate
$\langle \tau^{-1}_{\rm tot} \rangle = \langle \tau_{\rm ph}^{-1} \rangle  + \langle \tau_{i}^{-1} \rangle $. 
The dashed lines indicate the individual resistivity by impurity scattering $\rho_i$ and phonon scattering $\rho_{\rm ph}$. The difference of the red solid line from the black solid line shows the departure from Matthiessen's rule.
}
\label{fig:three}
\end{figure}

The Matthiessen's rule on the additivity of resistivities for different scattering mechanisms will not hold in most cases at finite temperatures. In general, the energy-averaged scattering rates do not add because the energy averaging is for $\tau$ rather than for $1/\tau$. We show that  the Matthiessen rule, i.e., $\rho_{\rm tot} = \rho_{\rm ph} + \rho_i$, where $\rho_{\rm tot}$ is the total resistivity contributed by impurities  $\rho_i$  and phonons  $\rho_{\rm ph}$, is simply not valid in the small angle twisted bilayer graphene.

In Fig.~\ref{fig:three} we show the calculated resistivity as a function of temperature for various Fermi velocities (twist angles), (a) $v_F=10^7$ cm/s ($\theta=1.24^{\circ}$), (b) $v_F=5\times 10^6$ cm/s ($\theta=1.14^{\circ}$), (c) $v_F=2.5\times 10^6$ cm/s ($\theta=1.09^{\circ}$), and (d) $v_F=2.1\times 10^6$ cm/s ($\theta=1.08^{\circ}$), considering two types of impurity scattering (screened long range impurity and unscreened short range impurity) and phonon scattering. The screened charged impurity density $n_i=5\times 10^{10}$ cm$^{-2}$ and the short range disorder strength $n_dV_0 = 0.72$ eV$^2$nm$^2$ are used. The deformation potential $D=15$ eV is used for the acoustic phonon scattering throughout this paper. By calculating the total resistivity arising from impurity scattering $\tau_i$ and phonon scattering $\tau_{\rm ph}$ we clearly show that $\rho_{\rm tot} > \rho_{\rm ph} + \rho_i$ for lower Fermi velocities. It is obvious from our results in Fig.~\ref{fig:three} 
that the Matthiessen's rule may fail badly for TBLG at low twist angles where the Fermi velocity (i.e., the twist angle) is small.
However, the differences are smaller for larger Fermi velocities (or larger angle TBLG).

\begin{figure}[t]
\vspace{10pt}%
\includegraphics[width=1.\linewidth]{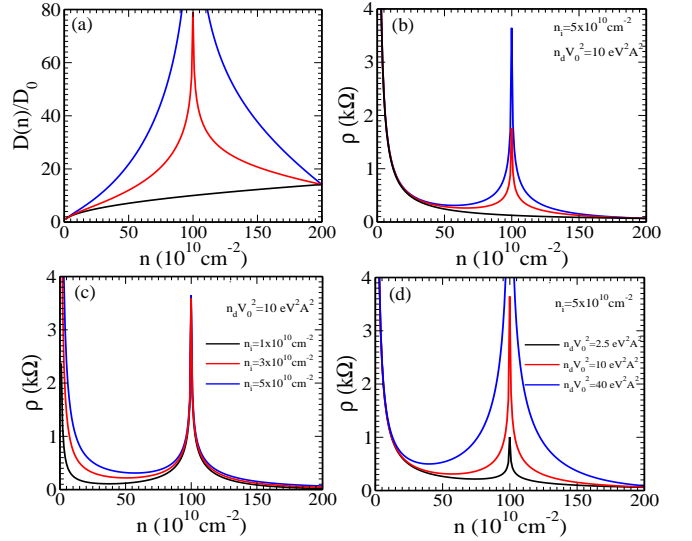}
\caption{(a) The density of states as a function of carrier density with $\alpha=0$ (black), 1 (red), and 3 (blue), and $n_0=10^{12}$ cm$^{-2}$. $v_F =10^7$ cm/s are used in this figure. 
(b) The calculated resistivity as a function of carrier density with the density of states corresponding to (a).  
(c) and (d) show the calculated resistivity for different combinations of charged disorder and neutral disorder with $\alpha = 3$. 
}
\label{fig:four}
\end{figure}

\begin{figure}[t]
\vspace{10pt}%
\includegraphics[width=1.\linewidth]{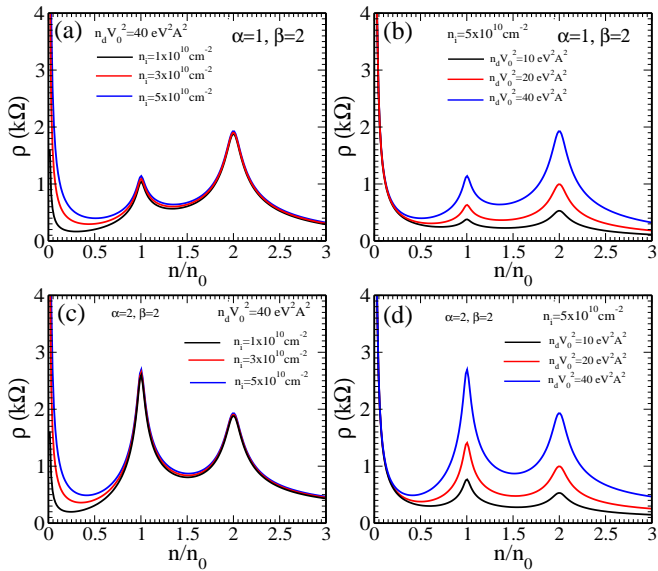}
\caption{ 
(a) and (b) Calculated resistivity for different combinations of charged disorder and neutral disorder with $\alpha = 1$ and $\beta=2$ in Eq.~(\ref{eq:dos2}). (c) and (d) show the calculated  resistivity with $\alpha=\beta=2$. 
}
\label{fig:five}
\end{figure}
 
\section{Inclusion of van Hove singularity} 

The above results are for low chemical potential with typical doping densities being around the Dirac point ($<2\times10^{12}$ cm$^{-2}$) so that the TBLG is less than quarter-filled.  At higher densities, as a vHS is approached, the theory must include the vHS in some manner.  Here, we demonstrate the vHS effect simply by considering various model density of states (DOS) as described below.  Our theory of transport in the presence of vHS using these simple model DOS should be taken as a zeroth order impurity scattering theory establishing the qualitative importance of vHS in determining TBLG transport at higher doping densities. We use the density of states having a logarithmic van Hove singularity at a density $n_0$ such as
\begin{equation}
D(n) = D_0\sqrt{n/n_0} \left \{ 1 + \frac{\alpha}{2} \left | \ln \left [ (1-n/n_0)^2 \right ] \right | \right \},
\label{eq:dos}
\end{equation}
where $D_0 = \sqrt{g n_0/\pi}/\hbar v_F$ and $\alpha$ is a parameter which controls the strength of the singularity of DOS at $n=n_0$.

In Fig.~\ref{fig:four} the resistivity, which is calculated with the density of states corresponding to Eq.~(\ref{eq:dos}), is shown as a function of carrier density. In Fig.~\ref{fig:four}(a) we show the density of states for three different $\alpha$=0, 1, 3 and for $n_0=10^{12}$ cm$^{-2}$.
Fig.~\ref{fig:four}(b) show the calculated resistivity for different $\alpha$ with a fixed charged impurity density $n_i = 5\times 10^{10}$ cm$^{-2}$ and a fixed neutral disorder strength $n_dV_0^2 = 10$ eV$^2$\AA$^2$. A resistivity peak appears at the singular point of DOS and the strength of the peak is strongly correlated to the singular feature of DOS. The resistivity for different combinations of charged impurity density and neutral disorder strength is shown in Fig.~\ref{fig:four}(c) and (d) for $\alpha = 3$. The calculated resistivity is significantly (weakly) affected by long range disorder at low carrier densities (near the singular point of DOS). However, as shown in Fig.~\ref{fig:four}(d) the calculated resistivity is weakly (significantly) affected by short range disorder at low carrier densities (near the singular point of DOS).

In Fig.~\ref{fig:five} the resistivity is calculated with the DOS having two singularities at $n_0$ and $2n_0$, i.e.,
\begin{eqnarray}
D(n) = D_0\sqrt{n/n_0}&   \{1  +  \frac{\alpha}{2} \left | \ln \left [ (1-\frac{n}{n_0})^2 + \eta \right ] \right |   \nonumber \\
& +  \frac{\beta}{2} \left | \ln \left [ (1-\frac{n}{2n_0})^2 + \eta \right ] \right |  \},
\label{eq:dos2}
\end{eqnarray}
where $\alpha$ and $\beta$ are constants determining the strength of the singularities and $\eta$ is introduced as a broadening to suppress the singular behaviors of DOS.
Figs.~\ref{fig:five}(a) and (b) show the calculated resistivity for different combinations of charged disorder and neutral disorder with $\alpha = 1$ and $\beta=2$, where the strength of the second peak in DOS  is two times larger than that of the first peak. 
The resistivity is also shown in Figs.~\ref{fig:five}(c)(d) for $\alpha=\beta=2$.
It is clear from these results that vHS have profound effects on TBLG transport.  In particular, Fig.~\ref{fig:five} shows that the resistivity in the presence of multiple vHS manifests a plateau-like almost-constant structure in between the vHS-induced resistivity maxima.  Such peak and plateau type resistivity in between various commensurate filling has actually been observed as a function of doping density in Ref.~[\onlinecite{six}], and we believe that vHS is the physical mechanism underlying these resistivity plateaus.
 
\section{Conclusion} 

We have theoretically calculated impurity scattering induced TBLG transport showing the profound effects of the velocity suppression and van Hove singularity in the moir\'e system.  In particular, the Matthiessen's rule is strongly violated in TBLG at small twist angles (where $v_F$ is low) and the presence of vHS produces resistivity peaks with plateau-like density-dependent resistivity in between the peaks.  
Our work, along with earlier work on phonon scattering establishes that TBLG has highly nontrivial and intriguing normal state transport properties in addition to having interesting collective ground states.

We mention that our minimal transport theory is motivated by the fact that at the current stage in the development of the subject, experimental transport data from different samples show considerable deviations so that focusing on a detailed modeling of the electronic structure in the transport theory is premature.  There is no question that subsequent theories will have to extend our minimal theory including many realistic effects neglected in our theory.  This is similar to the situation encountered in the studies of transport in simple monolayer graphene where early theories focused only on the linear chiral Dirac dispersion and the roles of short-range and long-range impurity scattering potentials without incorporating realistic band structure and realistic impurity scattering.\cite{eight,nine,ten}  These minimal theories of graphene transport turned out to explain much of the experimental data when the experiments eventually produced consistent results.\cite{eleven,twelve,thirteen,fourteen,fifteen}  Our hope is that our minimal theory for TBLG transport will serve a similar purpose providing the zeroth order understanding of ohmic transport.  All we have achieved in the current work is at best a semi-quantitative agreement with the experiment of Ref. [\onlinecite{three}], but future improvements should include realistic electronic structure information in order to make the theory more realistic and of quantitative validity.

Before concluding, it may be worthwhile to emphasize the many approximations and simplifications made in the current work so that future theories could take more realistic effects of experimental samples into account. The current work on impurity transport should be construed as a truly minimal model which includes the two most important and essential effects of TBLG band structure: flatband induced Fermi velocity suppression and low-energy van Hove singularities.  Our minimal model incorporates these two effects approximately, leaving out all other possible effects, which we believe to be less important. However, we include carrier chirality of the TBLG even though we neglect the details of the TBLG electronic structure. In addition, we consider only two types of impurity scattering in our theory: zero-range (by point defects) scattering and long-range (by charged impurities) Coulomb scattering. 
These two mechanisms are known to be\cite{eight,nine,ten,eleven,twelve,thirteen,fourteen,fifteen} the main operational impurity scattering mechanisms in regular monolayer and bilayer graphene, and as such, it makes sense to include these two mechanisms in the first theoretical work on impurity scattering effects in TBLG transport.  One serious problem in this context is that the details of the TBLG electronic structure are not yet well-established with the role of strain, substrates, higher bands, etc. are still being debated and analyzed in the literature.  In addition, little is known about the applicable disorder in TBLG (apart from possible long- and short-range disorder in the starting graphene layers, which our theory includes), and at this early stage of the subject it makes some sense to work within a highly simplified model which includes only the two essential features (i.e. long- and short-range impurity scattering) of TBLG impurity disorder.  

We also neglect the effects of intervalley scattering in the calculation. It is well know in graphene that the intervalley acoustic phonon scattering is negligible below room temperature \cite{seventeen} and comparable to the intravalley scattering above room temperature. For disorder potentials, it is rather general in 2D materials that intervalley impurity scatterings are strongly suppressed because the large momentum separation between the valleys makes the transitions between them difficult, as compared with the intravalley processes. 
More realistic future calculations including realistic band structures are necessary for future progress in our understanding of TBLG transport.

Future TBLG transport theories should include additional effects such as twist angle disorder \cite{new1}, which is specific to TBLG and does not exist in monolayer and bilayer graphene.  A strong motivation for us to keep our model minimal is also our desire to keep the theory analytical as much as possible so that our qualitative results and how they depend on the specific features of our model are manifestly clear.  At this early stage of the subject, it is of some advantage to focus on a minimal analytical model whose results are at best of qualitative validity.  We believe that our qualitative findings of the failure of the Matthiessen's rule and the plateau-like density-dependence of the resistivity would survive future improvements in transport calculations which may include more sophisticated and realistic effects of TBLG physics.

\section*{ ACKNOWLEDGMENTS}
This work is supported by the Laboratory for Physical Sciences. E.H.H. also acknowledges support from Basic Science Research Program No. 2017R1A2A2A05001403 of the National Research Foundation of Korea.

\begin{figure}[t]
\vspace{10pt}%
\includegraphics[width=.9\linewidth]{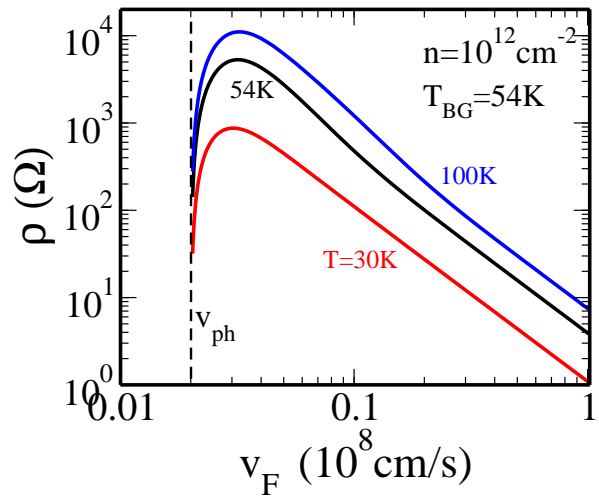}
\caption{
The calculated acoustic phonon-induced resistivity as a function of Fermi velocity for different temperatures and for an electron density $n=10^{12}cm^{-2}$. Here $T_{BG} = 54$K with the phonon velocity $v_{ph} = 2\times 10^{6}$ cm/s. The resistivity decreases as $v_F \rightarrow v_{ph}$ due to the strong phase-space restriction. 
}
\label{fig:append1}
\end{figure}

\appendix
\section{}

In Eq.~(\ref{eq2_time}) the transport relaxation time  $\tau_{ph}$ is calculated by considering  
the deformation potential coupled acoustic phonon mode.
Taking $\vk$ and $\vk'$ to denote the electron wave vectors before and after scattering by a phonon, respectively, the energy dependent relaxation time ($\tau_{ph}$) is defined by
\begin{equation}
\frac{1}{\tau_{ph}(\ve_{\vk})} = \sum_{\vk'}(1-\cos\theta_{\vk \vk'}) W_{\vk
  \vk'}\frac{1 - f(\ve_{\vk})}{1-f(\ve_{\vk'})}
\label{tau_3d}
\end{equation}
where $\theta_{\vk \vk'}$ is the angle between $\vk$ and $\vk'$, and
$W_{\vk \vk'}$ is the transition rate from the state with 
momentum $\vk$ to $\vk'$ state. 
When we consider the relaxation time due to deformation potential (DP) coupled acoustic phonon mode, 
then the transition rate has the form
\begin{equation}
W_{\vk \vk'}=\frac{2\pi}{\hbar}|C(\vq)|^2
\Delta(\varepsilon,\varepsilon')   
\label{wkkp}
\end{equation}
where $\vq = \vk-\vk'$ and $|C(\vq)|^2$ is the matrix element for scattering by acoustic phonon. The matrix element $|C(\vq)|^2$ for the deformation potential is given by
\begin{equation}
|C(\vq)|^2 = \frac{D^2\hbar q}{2\rho_0 v_{s}},
\end{equation}
where $D$ is the deformation potential and $\rho_0$ is the mass density. 
In Eq.~(\ref{wkkp}) the factor $\Delta(\ve,\ve')$ is given by
\begin{equation}
\Delta(\ve,\ve') = N_q \delta(\ve-\ve'+\hbar\omega_{\vq}) + (N_q + 1)
\delta(\ve-\ve'-\hbar\omega_{\vq}),
\label{delta}
\end{equation}
where  $\omega_{\vq}=v_{ph} \vq$ is the acoustic phonon energy with $v_{ph}$ being the phonon velocity,  $\ve = \vek$, $\ve' = \ve_{k'}$,  and
$N_q$ is the phonon occupation number
$N_q = [{\exp(\beta \omega_{\vq}) -1}]^{-1},$
where $\beta = k_B T$.
The first (second) term in Eq.~(\ref{delta}) corresponds to the
absorption (emission) of an acoustic phonon of wave vector $\vq = \vk-\vk'$.
Note that the matrix element $|C(\vq)|^2$ is independent of the phonon
occupation numbers.

The scattering of electrons from acoustic phonons can be considered quasi elastically when $\hbar \omega_{\vq} \ll E_F$, where $E_F$ is the Fermi energy. In this case the relaxation time is calculated to be
\begin{equation}
\frac{1}{\tau_{ph}(\ve_{\vk})} =
\frac{1}{\hbar^2}\frac{\ve_{\vk}}{4v_F^2}\frac{D^2}{\rho v_{ph}^2}
k_BT.
\end{equation}
Thus, in the equipartition regime ($\hbar \omega_{\vq} \ll k_B T$) we have the conductivity arising from phonon scattering 
\begin{equation}
\sigma_{ph}(n,T) = \frac{\sigma_0(n,T)}{e^{-\mu(T)/k_BT} + 1},
\end{equation} 
where
\begin{equation}
\sigma_0(n,T) = \frac{e^2}{h} \frac{2g(\hbar v_F)^2\rho_m v_{ph}^2}{D^2} \frac{1}{k_BT}.
\end{equation} 
In the low temperature limit, $T\ll T_F$, $\sigma_{ph}(n,T) \rightarrow \sigma_0(n,T)$, but in the high temperature (non-degenerate) limit, $T\gg T_F$, we have $\sigma_{ph}(n,T) \rightarrow \sigma_0(n,T)/2$.

As the Fermi velocity approaches phonon velocity ($v_F \rightarrow v_{ph}$) the available scattering process is severely restricted.
The electron-phonon scattering  must satisfy the following energy-momentum conservation laws for the scattering process $\vk \rightarrow \vk'$
\begin{eqnarray}
\ve' & = &\ve \pm \hbar \omega_q, \nonumber \\
q & = & \left [ k^2 + k'^2 -2 k k' \cos \theta \right]^{1/2}.
\end{eqnarray}
When $v_F \gg v_{ph}$ we may use the quasielastic condition $\varepsilon' = \varepsilon$. However, for $v_F \sim v_{ph}$ the scattering angle $\theta$ is very restricted and only small angles (forward scattering) satisfy the energy-momentum conservation laws. As shown in Fig.~\ref{fig:append1}, for $v_F \sim v_{ph}$ $\rho(v_F)$ do  not follow the relation $\rho(v_F) \propto v_F^{-2}$ anymore, and the resistivity is strongly reduced near $v_F$. Furthermore, when $v_{ph} \ge v_F$ there are no allowed phase space for the scattering process, which gives rise to the zero resistivity. In this case, the Umklapp process may contribute to the resistivity, but this process is not included in the calculation.

\end{document}